\documentclass[10pt,twocolumn,twoside,english]{IEEEtran}
\usepackage[T1]{fontenc}
\usepackage{mathrsfs}
\usepackage{bm}
\usepackage{amsmath}
\usepackage{amsthm}
\usepackage{amssymb}
\usepackage{graphicx}
 \PassOptionsToPackage{normalem}{ulem}

\makeatletter

\theoremstyle{plain}

  \theoremstyle{plain}
  
  \theoremstyle{plain}
  
  \theoremstyle{plain}
   
  \theoremstyle{remark}

\theoremstyle{assumption}
    \theoremstyle{proposition}
  \newtheorem{proposition}{\protect\propositionname}
  
\theoremstyle{algorithm}  
  
\usepackage{amsmath}
\usepackage{amssymb}
\usepackage{graphicx,psfrag,cite,subfigure}
\usepackage[table]{xcolor}

\setkeys{Gin}{width=1.0\columnwidth}

\makeatother

\usepackage{babel}
  \providecommand{\definitionname}{Definition}
  \providecommand{\lemmaname}{Lemma}
  \providecommand{\propositionname}{Proposition}
  \providecommand{\remarkname}{Remark}
\providecommand{\theoremname}{Theorem}
\providecommand{\conjecturename}{Conjecture}
\providecommand{\assumptionname}{Assumption}
\providecommand{\algorithmname}{Algorithm}

\begin{document}

 \title{Optimal deception attack on
networked vehicular cyber physical systems
 \thanks{Moulik Choraria and Arpan Chattopadhyay are with the Department of Electrical Engineering, Indian Institure of Technology, Delhi. Email: moulik.choraria@gmail.com, arpanc@ee.iitd.ac.in. Urbashi Mitra is with the Department of Electrical Engineering, University of Southern California. Email: ubli@usc.edu. Erik Strom is with the Department of Signals and Systems, Chalmers University, Sweden. Email: erik.strom@chalmers.se.}
 \thanks{This work was supported by the faculty seed grant and professional development allowance (PDA) of IIT Delhi, and one or more   of the following grants:
ONR N00014-15-1-2550, NSF CCF-1718560,NSF CCF-1410009,
NSF CPS-1446901, NSF CCF-1817200, and ARO 74745LSMUR.
}
}

\author{
Moulik Choraria, Arpan Chattopadhyay,  Urbashi Mitra, Erik Strom
}

\maketitle
%
%



\ifdefined\SINGLECOLUMN
	\setkeys{Gin}{width=0.5\columnwidth}
	\newcommand{\figfontsize}{\footnotesize} 
\else
	\setkeys{Gin}{width=1.0\columnwidth}
	\newcommand{\figfontsize}{\normalsize} 
\fi

\begin{abstract} 
Herein, design of false data injection attack on a distributed cyber-physical system is considered. A stochastic process with linear dynamics and Gaussian noise is measured by multiple agent nodes, each equipped with multiple sensors. The agent nodes form a multi-hop network among themselves. Each agent node computes an estimate of the process by using its sensor observation and messages obtained from neighbouring nodes, via Kalman-consensus filtering. An external attacker, capable of arbitrarily manipulating the sensor observations of some or all agent nodes, injects errors into those sensor observations. The goal of the attacker is to steer the estimates at the agent nodes as close as possible to a pre-specified value, while respecting a constraint on the attack detection probability. To this end, a constrained optimization problem is formulated to find the optimal parameter values of a certain class of linear attacks. The parameters of linear attack are learnt on-line via a combination of stochastic approximation and online stochastic gradient descent. Numerical results demonstrate the efficacy of the  attack.
\end{abstract}

\section{Introduction}\label{section:introduction}
In recent times, there have been significant interest in designing cyber-physical systems (CPS) that  combine the cyber world and the physical world via seamless integration of sensing, computation, communication, control and learning. CPS has widespread applications such as networked  monitoring and control of industrial processes, disaster management, smart grids, intelligent transportation systems, etc. These applications critically depend on estimation of a physical process via multiple sensors over a wireless network. However, increasing use of wireless networks in sharing the sensed data has rendered the sensors vulnerable to various cyber-attacks.  In this paper, we focus on {\em false data injection} (FDI) attacks which is an integrity or deception attack where the attacker modifies the information flowing through the network \cite{mo2009secure, mo2014detecting}, in contrast  to a {\em denial-of-service} attack where the attacker blocks system resources ({\em e.g.},  wireless jamming attack  \cite{guan2018distributed}). In FDI, the attacker either breaks the cryptography of the data packets or physically manipulates the sensors ({\em e.g.}, putting a heater near a temperature sensor). 

\subsection{Related Literature}
The cyber-physical systems either need to compute the process estimate in a remote estimator ({\em centralized} case), or often multiple nodes or components of the system need to estimate the same process over time via sensor observations and the information shared over a network ({\em distributed} case). 
The problem of FDI attack design and its detection has received significant  attention in recent times; attack design:  conditions for    undetectable FDI  attack \cite{chen2017optimal}, design of a linear deception attack scheme to fool the popular $\chi^2$ detector (see \cite{guo2017optimal}), optimal attack design for noiseless systems \cite{wu2018optimal}. The paper  \cite{chen2016cyber} designs an optimal attack to steer the  state of a control system to a desired target under a  constraint on the attack detection probability. On the other hand, attempts on attack detection includes centralized (and decentralized as well)     schemes for {\em noiseless} systems  \cite{pasqualetti2013attack}, coding of sensor output along with $\chi^2$ detector   \cite{miao2017coding},  comparing the sensor observations with those coming from from a few {\em known safe} sensors  \cite{li2017detection}, and the attack detection and secure estimation schemes based on innovation vectors in \cite{mishra2017secure}.  Attempts on attack-resilient state estimation include: \cite{pajic2017attack} for  {\em bounded} noise, \cite{chattopadhyay2018attack, chattopadhyay2018secure, chattopadhyay2019security} for adaptive filter design using stochastic approximation,  \cite{liu2017dynamic} that uses  sparsity models to characterize the switching location attack in a {\em noiseless} linear system and    state recovery constraints for various attack modes. FDI attack and its mitigation  in power  systems are addressed in \cite{manandhar2014detection, liang2017review, hu2017secure}.  Attack-resilient state estimation and control in noiseless systems are discussed in  \cite{nakahira2018attack} and \cite{fawzi2014secure}. Performance bound of stealthy attack in a single sensor-remote estimator system using Kalman filter was characterized in \cite{bai2017kalman}.

However, there have been very few attempts for attack design and mitigation in distributed CPS, except \cite{guan2017distributed} for attack detection and secure estimation, \cite{satchidanandan2016dynamic} for attack detection in networked control system using a certain {\em dynamic watermarking} strategy, and \cite{dorfler2011distributed} for attack detection in power systems. To our knowledge, there has been no attempt to theoretically design an attack strategy in distributed CPS. 

\subsection{Our Contribution}
In light of the above issues, our contributions in this paper are the following:

\begin{enumerate}
    \item Under the Kalman-consensus filter (KCF, see \cite{saber09kalman-consensus-optimality-stability}) for distributed estimation, we design a novel attack scheme that steers the estimates in all estimators towards a target value, while respecting a constraint on the attack detection probability under the popular $\chi^2$ detector adapted to the distributed setting. The attack scheme is reminiscent of the popular linear attack scheme \cite{guo2017optimal}, but the novelty lies in online learning of the parameters in the attack algorithm via simultaneous perturbation stochastic approximation (SPSA, see \cite{spall92original-SPSA}). The attack algorithm, unlike the linear attack scheme of  \cite{guo2017optimal}, uses a non-zero mean Gaussian perturbation to modify the observation made at a node, and this non-zero mean is an affine function of the process estimate at a node. The optimization problem is cast as an online  optimization problem, where SPSA is used for online stochastic gradient descent (see \cite[Chapter~$3$]{hazan2016introduction}). 
    \item The constraint on attack detection probability is met by updating a Lagrange multiplier via stochastic approximation at a slower timescale.
    \item Though the proposed algorithm involves on-line parameter learning, it can be used off-line to optimize the attack parameters which can then be used in real CPS.
\end{enumerate}
 
\subsection{Organization}
The rest of the paper is organized as follows. System model and the necessary background related to the problem are provided in Section~\ref{section:background}. The attack design algorithm is developed in Section~\ref{section:attack-scheme}, and its  performance in evaluated numerically in Section~\ref{section:numerical-work}. Conclusions are made in Section~\ref{section:conclusion}.

\section{System Model and Background}\label{section:background}
In this paper, bold capital letters, bold small letters   and capital letters with caligraphic font  will denote matrices, vectors and sets respectively.

\subsection{Sensing and  estimation model}\label{subsection:sensing-model}
\begin{figure}[t!]
 \begin{centering}
 \begin{center}
 \includegraphics[height=6cm, width=7cm]{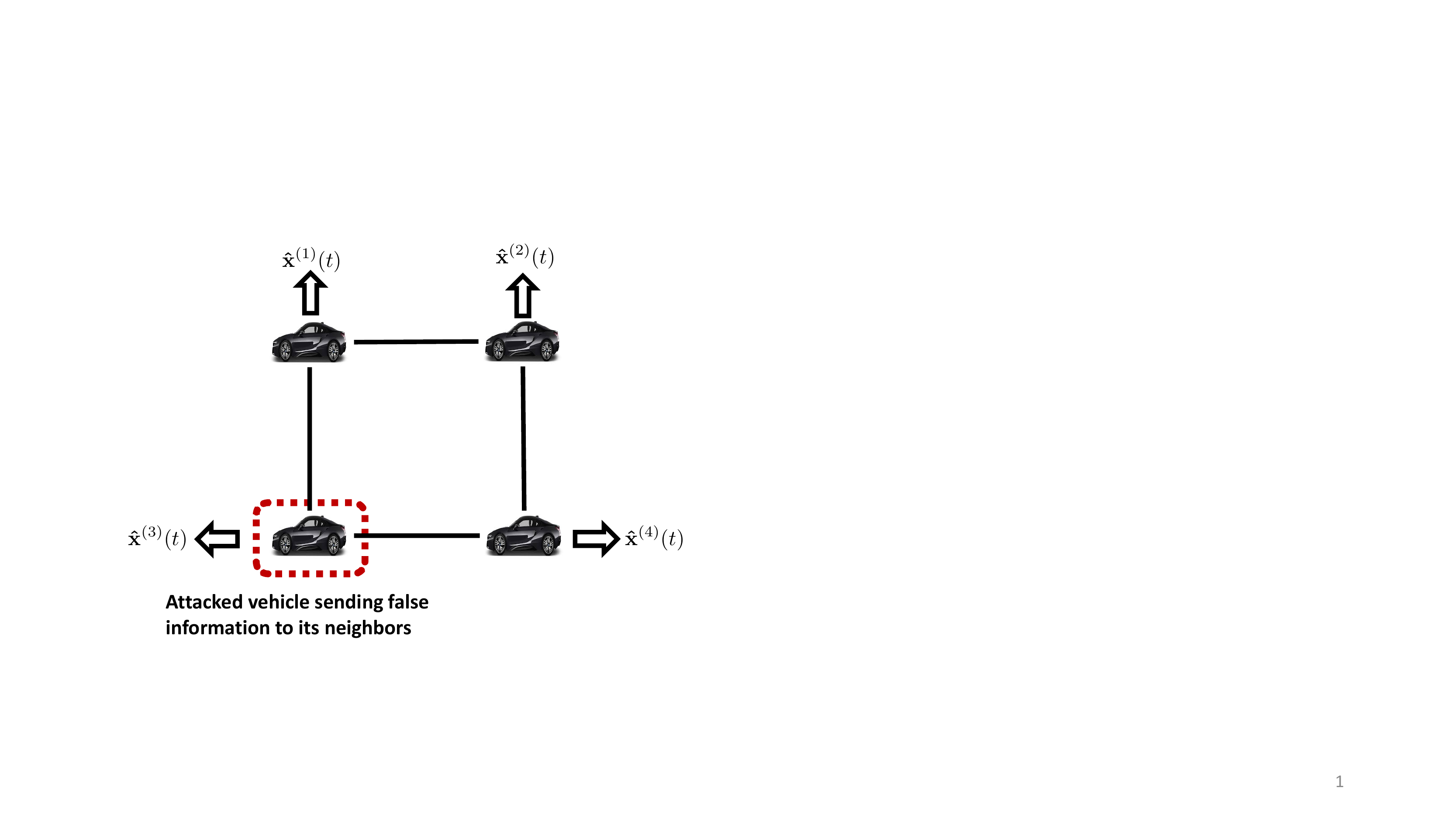}
 \end{center}
 \end{centering}
 \vspace{-5mm}
 \caption{False data injection attack in remote estimation.}
 \label{fig:distributed-attack}
 \vspace{-5mm}
 \end{figure}

We consider a connected, multihop wireless network (see Figure~\ref{fig:distributed-attack}) of $N$ agent nodes denoted by $\mathcal{N}\doteq\{1,2,\cdots,N\}$. The set of neighbouring nodes of node~$k$ is denoted by $\mathcal{N}_k$, and let $N_k \doteq |\mathcal{N}_k|$.  There is a discrete-time stochastic process $\{\bm{x}(t)\}_{t \geq 0}$ (where $\bm{x}(t) \in \mathbb{R}^q$) which is a linear Gaussian process evolving as follows:
\begin{equation}\label{eqn:process-equation}
\bm{x}(t+1)=\bm{A} \bm{x}(t)+\underbrace{\bm{w}(t)}_{\sim \mathcal{N}(\bm{0}, \bm{Q})} 
\end{equation}
where $\bm{w}(t)$ is  zero-mean i.i.d. Gaussian noise. 

Each agent node is equipped with one or more sensors who make some observation about the process. The  vector observation received at node~$k$ at time~$t$  is given by:
\begin{equation}
\bm{y}_k(t)=\bm{H}_k \bm{x}(t)+\underbrace{\bm{v}_k(t)}_{\sim \mathcal{N}(\bm{0}, \bm{R}_k)},  \label{eqn:observation-equation}
\end{equation}
where $\bm{H}_k$ is a matrix of appropriate dimension and $\bm{v}_k(t)$ is a Gaussian observation noise which is independent across sensors and i.i.d. across $t$. 

At time $t$, each agent node~$k \in \mathcal{N}$ declares an estimate $\bm{\hat{x} }^{(k)}(t)$ using Kalman consensus filtering (KCF, see \cite{saber09kalman-consensus-optimality-stability}) which involves the following sequence of steps:
\begin{enumerate}
\item Node~$k$ computes an intermediate estimate $\bar{\bm{x}}^{(k)}(t)=\bm{A}\hat{\bm{x}}^{(k)}(t-1)$.
\item Node~$k$ broadcasts $\bar{\bm{x}}^{(k)}(t)$ to all $j \in \mathcal{N}_k$. 
\item Node~$k$ computes its final estimate of the process as:
\begin{eqnarray}\label{eqn:KCF-equation}
 \hat{\bm{x}}^{(k)}(t)&=&\bar{\bm{x}}^{(k)}(t)+\bm{G}_k (\bm{y}_k(t)-\bm{H}_k \bar{\bm{x}}^{(k)}(t)) \nonumber\\
 && +\bm{C}_k \sum_{j \in \mathcal{N}_k} (\bar{\bm{x}}^{(j)}(t)-\bar{\bm{x}}^{(k)}(t))
\end{eqnarray}
\end{enumerate}
Here $\bm{G}_k$ and $\bm{C}_k$ are the Kalman and consensus gain matrices used by node~$k$, respectively.

\subsection{False data injection (FDI) attack}\label{subsection:FDI-attack}
At time~$t$, sensors associated to any  subset of nodes $\mathcal{A}_t \subset \mathcal{N}$ can be under attack. A node~$k \in \mathcal{A}_t$ receives an  observation:
\begin{equation}\label{eqn:attack-equation}
\tilde{\bm{y}}_k(t)=\bm{H}_k \bm{x}(t)+\bm{e}_k (t)+\bm{v}_k(t),  
\end{equation}
where $\bm{e}_k(t)$ is the error injected by the attacker. The attacker seeks to insert the error sequence $\{\bm{e}_k(t): k \in \mathcal{A}_t\}_{t \geq 0}$   in order to introduce error in the estimation. 
If $\mathcal{A}_t=\mathcal{A}$ for all $t$, then the attack is called a {\em static attack}, otherwise the attack is called a {\em switching location attack}. {\em We will consider only static attack in this paper.} We assume that the attacker can observe $\hat{\bm{x}}^{(k)}(t)$ for all $1 \leq k \leq N$ once they are computed by the agent nodes.

\subsection{The $\chi^2$ detector} \label{subsection:chi-square-detector}
Let us define the innovation vector at node~$k$ by $\bm{z}_k(t):=\bm{y}_k(t)-\bm{H}_k \bm{A} \hat{\bm{x}}^{(k)}(t-1)$.  Let us assume that, under no attack, 
$\{\bm{z}_k(t)\}_{t \geq 0}$ reaches its steady-state distribution 
 $N(\bm{0}, \bm{\Sigma}_k)$. Under a possible attack, a  standard technique (see \cite{guo2017optimal}, \cite{li2017detection}) to detect any anomaly in $\{\bm{z}_t\}_{t \geq 0}$ is the $\chi^2$ detector, which tests whether the innovation vector follows the desired Gaussian distribution. The detector {\em at each agent node} observes the innovation sequence over a pre-specified window of $J$ time-slots, and declares   an attack at time $\tau$   if  
$\sum_{t=\tau-J+1}^{\tau} \bm{z}_k(t)' \bm{\Sigma}_k^{-1} \bm{z}_k(t) \geq \eta$,  
where $\eta$ is a threshold  which can be adjusted to control the false alarm probability. The covariance matrix $\bm{\Sigma}_k$ can be computed from standard results on KCF as in \cite{saber09kalman-consensus-optimality-stability}.

The authors of  \cite{guo2017optimal} proposed a linear injection attack to fool the $\chi^2$ detector in a centralized, remote estimation setting. Motivated by  \cite{guo2017optimal}, we also propose a linear attack, where,  at time $t$, the    sensor(s) associated with any node~$k \in \mathcal{A}$   modifies the innovation vector as $\bm{\tilde{z}_k}(t)=\bm{T}_k \bm{z}_k(t)+\bm{b}_k(t)$, where $\bm{T}_k$ is a square matrix and $\bm{b}_k(t) \sim N (\bm{\mu}_k( \bm{\theta}^{(k)}(t-1)),\bm{S}_k)$ is  independent Gaussian. The bias term $\bm{\mu}_k(\bm{\theta}^{(k)}(t-1))$ is assumed to take a linear form $\bm{\mu}_k(\bm{\theta}^{(k)}(t-1))=\bm{M}_k \bm{\theta}^{(k)}(t-1)+\bm{d}_k$ for suitable matrices and vectors $\bm{M}_k$ and $\bm{d}_k$. This is equivalent to modifying the observation vector to $\bm{\tilde{y}_k}(t)$. If $\{\bm{T}_k, \bm{S}_k, \bm{M}_k, \bm{d}_k\}_{1 \leq k \leq N}$ is constant over time~$t$, the attack is called stationary, else non-stationary.

\subsection{The optimization problem}\label{subsection:optimization-problem}
The attacker seeks to steer the estimate as close as possible to some pre-defined value $\bm{x}^*$, while keeping the attack detection probability per unit time under some constraint value $\alpha$. Note that, the probability of attack detection per unit time slot can be upper bounded as:
\begin{eqnarray}
  P_d &\leq& \limsup_{T \rightarrow \infty} \frac{1}{T+1} \sum_{\tau=0}^T \sum_{k=1}^N \mathbb{P}(\sum_{t=\tau-J+1}^{\tau} \tilde{\bm{z}}_k(t)' \bm{\Sigma}_k^{-1} \tilde{\bm{z}}_k(t) \geq \eta) \nonumber\\
  &\leq & \limsup_{T \rightarrow \infty} \frac{1}{T+1} \sum_{\tau=0}^T \sum_{k=1}^N \frac{\mathbb{E}(\sum_{t=\tau-J+1}^{\tau} \tilde{\bm{z}}_k(t)' \bm{\Sigma}_k^{-1} \tilde{\bm{z}}_k(t))}{\eta} \nonumber\\
  &= & \frac{J}{\eta} \limsup_{T \rightarrow \infty} \frac{1}{T+1} \sum_{\tau=0}^T \sum_{k=1}^N \mathbb{E} (\tilde{\bm{z}}_k(t)' \bm{\Sigma}_k^{-1} \tilde{\bm{z}}_k(t) )
\end{eqnarray}
where the first and second inequalities come from union bound and Markov inequality, respectively. Hence, the attacker seeks to solve the following constrained optimization problem:

\small
\begin{align}\label{eqn:constrained-optimization-problem} 
 && \min_{ \{\bm{T}_k, \bm{S_k}, \bm{M}_k, \bm{d}_k \}_{k=1}^N } \limsup_{T \rightarrow \infty} \frac{1}{T+1} \sum_{t=0}^T \sum_{k=1}^N \mathbb{E}||\hat{\bm{x}}^{(k)}(t)-\bm{x}^*||^2    \nonumber\\
 &\text{s.t. }& \limsup_{T \rightarrow \infty} \frac{1}{T+1} \sum_{t=0}^T \sum_{k=1}^N \mathbb{E} ( \tilde{\bm{z}}_k(t)' \bm{\Sigma}_k^{-1} \tilde{\bm{z}}_k(t) ) \leq \frac{\alpha \eta}{J} \tag{CP}
\end{align}
\normalsize

This problem can be relaxed by a Lagrange multiplier $\lambda$ to obtain the following unconstrained optimization problem:
\small
\begin{align}\label{eqn:unconstrained-optimization-problem}  
 \min_{ \{\bm{T}_k, \bm{S}_k,  \bm{M}_k, \bm{d}_k\}_{k=1}^N } && \limsup_{T \rightarrow \infty} \frac{1}{T+1} \sum_{t=0}^T \sum_{k=1}^N \mathbb{E}(||\hat{\bm{x}}^{(k)}(t)-\bm{x}^*||^2  \nonumber\\
 && +\lambda \tilde{\bm{z}}_k(t)' \bm{\Sigma}_k^{-1} \tilde{\bm{z}}_k(t) ) \tag{UP}
\end{align}
\normalsize

The following standard result tells us how to choose $\lambda$.
\begin{proposition}\label{proposition:choice-of-lambda}
 Let us consider \eqref{eqn:constrained-optimization-problem} and its relaxed version \eqref{eqn:unconstrained-optimization-problem}. If there exists a $\lambda^* \geq 0$ and matrices $\{\bm{T}_k^*, \bm{S}_k^*, \bm{M}_k^*, \bm{d}_k^*\}_{k=1}^N$ such that (i) $\{\bm{T}_k^*, \bm{S}_k^*, \bm{M}_k^*, \bm{d}_k^*\}_{k=1}^N$ is the optimal solution of \eqref{eqn:unconstrained-optimization-problem} under $\lambda=\lambda^*$, and 
 (ii) the tuple $(\{\bm{T}_k^*, \bm{S}_k^*, \bm{M}_k^*, \bm{d}_k^*\}_{k=1}^N,\lambda^*)$ satisfies the constraint in \eqref{eqn:constrained-optimization-problem} with equality, then $(\{\bm{T}_k^*, \bm{S}_k^*, \bm{M}_k^*, \bm{d}_k^*\}_{k=1}^N,\lambda^*)$ is an optimal solution for \eqref{eqn:constrained-optimization-problem} as well.
\end{proposition}

\section{Attack design}\label{section:attack-scheme}
We first analytically characterize the dynamics of the deviation 
$(\hat{\bm{x}}^{(k)}(t)-\bm{x}^*)$ in presence of linear attack, which will be used in developing the attack design algorithm later.

Our proposed algorithm maintains iterates $\{\bm{T}_k(t), \bm{U}_k(t), \bm{M}_k(t), \bm{d}_k(t)\}_{1 \leq k \leq N}$ for $\{\bm{T}_k, \bm{U}_k, \bm{M}_k, \bm{d}_k\}_{1 \leq k \leq N}$, where $\bm{U}_k' \bm{U}_k \doteq \bm{S}_k$. Let 
\begin{eqnarray}
 \mathcal{F}_{\tau} &\doteq& \sigma (\{\hat{\bm{x}}^{(k)}(t), \bm{y}_k(t), \bm{T}_k(t), \bm{U}_k(t), \bm{M}_k(t), \bm{d}_k(t),  \nonumber\\
&&\bm{b}_k(t)\}_{1 \leq k \leq N}, \lambda(t): 1 \leq  t \leq \tau  )
\end{eqnarray}
be a sigma algebra; this is the information  available to the attacker at time $(\tau+1)$ before   a new attack.

\subsection{Dynamics of deviation from target estimate}\label{subsection:dynamics-of-deviation}

Let us assume for the sake of analysis that the attacker uses constant $\bm{T}_k$ and $\bm{S}_k$ respectively, for all $k \in \{1,2,\cdots,N\}$. Under this attack:

\footnotesize
\begin{eqnarray}
 &&\hat{\bm{x}}^{(k)}(t) \nonumber\\
 &=& \bm{A} \hat{\bm{x}}^{(k)}(t-1)+ \bm{G}_k \tilde{\bm{z}}_k(t)
+ \bm{C}_k \sum_{j \in \mathcal{N}_k} (\bar{\bm{x}}^{(j)}(t)-\bar{\bm{x}}^{(k)}(t)) \nonumber\\
 &=& \bm{A} \hat{\bm{x}}^{(k)}(t-1)+ \bm{G}_k (\bm{T}_k(\bm{y}_k(t)-\bm{H}_k \bm{A} \hat{\bm{x}}^{(k)}(t-1))+\bm{b}_k(t)) \nonumber\\
 &+& \bm{C}_k \bm{A} \sum_{j \in \mathcal{N}_k} (\hat{\bm{x}}^{(j)}(t-1)-\hat{\bm{x}}^{(k)}(t-1)) 
\end{eqnarray}
\normalsize

Let us define $\bm{\theta}^{(k)}(t) \doteq \hat{\bm{x}}^{(k)}(t)-\bm{x}^*$. Let  $\tilde{\bm{\phi}}(t) \doteq (\hat{\bm{x}}(t)-\bm{x}(t))$, where 
$\hat{\bm{x}}(t) \doteq \mathbb{E}(\bm{x}(t)|\{ \bm{y}_k(\tau)\}_{1 \leq k \leq N, \tau \leq t})=\mathbb{E}(\bm{x}(t)|\mathcal{F}_{t})$ which can be computed by a standard Kalman filter. Let $\tilde{\bm{\phi}}(t) \sim \mathcal{N}(\bm{0}, \bm{R}(t))$, whose distribution can be computed by a standard Kalman filter. Hence, conditioned on $\mathcal{F}_t$, the distribution of 
$\bm{\phi}(t) \doteq (\bm{x}(t)-\bm{x}^*)$ is $\mathcal{N}(\hat{\bm{x}}(t)-\bm{x}^*, \bm{R}(t))$.

\footnotesize
\begin{eqnarray}
 &&\bm{\theta}^{(k)}(t) \nonumber\\
 &=& (\bm{A}-\bm{G}_k \bm{T}_k \bm{H}_k \bm{A}) \hat{\bm{x}}^{(k)}(t-1) \nonumber\\
 && + \bm{G}_k \bm{T}_k \underbrace{\bm{y}_k(t)}_{\doteq \bm{H}_k \bm{A}\bm{x}(t-1)+ \bm{H}_k \bm{w}(t-1)+\bm{v}_k(t)}+ \bm{G}_k \bm{b}_k(t)\nonumber\\
&& + \bm{C}_k \bm{A} \sum_{j \in \mathcal{N}_k} (\hat{\bm{x}}^{(j)}(t-1)-\hat{\bm{x}}^{(k)}(t-1))-\bm{x}^*\nonumber\\
 &=& (\bm{A}-\bm{G}_k \bm{T}_k \bm{H}_k \bm{A}) \bm{\theta}^{(k)}(t-1)+ \bm{G}_k \bm{T}_k \bm{H}_k \bm{A}\bm{\phi}(t-1) \nonumber\\
 && + \bm{C}_k \bm{A} \sum_{j \in \mathcal{N}_k} (\bm{\theta}^{(j)}(t-1)-\bm{\theta}^{(k)}(t-1))-(\bm{I}-\bm{A})\bm{x}^* \nonumber\\
 && + \bm{G}_k \bm{T}_k \bm{H}_k \bm{w}(t-1) + \bm{G}_k \bm{b}_k(t) + \bm{G}_k \bm{T}_k \bm{v}_k(t) \nonumber\\
  &=& (\bm{A}-\bm{G}_k \bm{T}_k \bm{H}_k \bm{A}-N_k \bm{C}_k \bm{A}) \bm{\theta}^{(k)}(t-1) \nonumber\\
 && + \bm{C}_k \bm{A} \sum_{j \in \mathcal{N}_k} \bm{\theta}^{(j)}(t-1)-(\bm{I}-\bm{A})\bm{x}^* + \bm{G}_k \bm{T}_k \bm{H}_k \bm{A}\bm{\phi}(t-1) \nonumber\\
 && + \bm{G}_k \bm{T}_k \bm{H}_k \bm{w}(t-1) + \bm{G}_k \bm{b}_k(t) + \bm{G}_k \bm{T}_k \bm{v}_k(t) \label{eqn:theta-evolution}
\end{eqnarray}
\normalsize

\begin{figure*}[t!]
 
  \footnotesize
 \begin{eqnarray}\label{eqn:variance-of-theta}
  \mathbb{E}(||\bm{\theta}^{(k)}(t)||^2| \mathcal{F}_{t-1})
  &=& ||(\bm{A}-\bm{G}_k \bm{T}_k \bm{H}_k \bm{A}-N_k \bm{C}_k \bm{A})\bm{\theta}^{(k)}(t-1)+ \bm{C}_k \bm{A} \sum_{j \in \mathcal{N}_k} \bm{\theta}^{(j)}(t-1)-(\bm{I}-\bm{A})\bm{x}^*+ \bm{G}_k (\bm{M}_k \bm{\theta}^{(k)}(t-1)+\bm{d}_k)||^2 \nonumber\\
  &&+ \mbox{Tr} ( \bm{G}_k \bm{T}_k \bm{H}_k \bm{Q} \bm{H}_k' \bm{T}_k' \bm{G}_k' + \bm{G}_k \bm{S}_k \bm{G}_k' + \bm{G}_k \bm{T}_k \bm{R}_k \bm{T}_k' \bm{G}_k' ) \nonumber\\
  && + 2 \bigg((\bm{A}-\bm{G}_k \bm{T}_k \bm{H}_k \bm{A}-N_k \bm{C}_k \bm{A})\bm{\theta}^{(k)}(t-1)+ \bm{C}_k \bm{A} \sum_{j \in \mathcal{N}_k} \bm{\theta}^{(j)}(t-1)-(\bm{I}-\bm{A})\bm{x}^* + \bm{G}_k (\bm{M}_k \bm{\theta}^{(k)}(t-1)+\bm{d}_k)) \bigg)' \nonumber\\
  && \bm{G}_k \bm{T}_k \bm{H}_k \bm{A} 
  \underbrace{\mathbb{E}( \bm{\phi}(t-1) |\mathcal{F}_{t-1})}_{=\hat{\bm{x}}(t-1)-\bm{x}^*}  + \underbrace{ \mathbb{E}(|| \bm{G}_k \bm{T}_k \bm{H}_k \bm{A}\bm{\phi}(t-1)||^2 |\mathcal{F}_{t-1}) }_{=\mbox{Tr}\bigg( \bm{G}_k \bm{T}_k \bm{H}_k \bm{A} \bigg(\bm{R}(t-1)+(\hat{\bm{x}}(t-1)-\bm{x}^*)(\hat{\bm{x}}(t-1)-\bm{x}^*)' \bigg) \bm{A}'\bm{H}_k' \bm{T}_k' \bm{G}_k' \bigg)}
  \end{eqnarray}
 
  \begin{eqnarray}\label{eqn:variance-of-z}
   \mathbb{E}(\tilde{\bm{z}}_k(t)' \bm{\Sigma}_k^{-1} \tilde{\bm{z}}_k(t) | \mathcal{F}_{t-1}) 
  &=& \mbox{Tr} \bigg( \bm{\Sigma}_k^{-\frac{1}{2}} \bigg( \bm{T}_k \bm{H}_k \bm{Q} \bm{H}_k' \bm{T}_k'+ \bm{T}_k \bm{R}_k \bm{T}_k' + \bm{S}_k + \bm{T}_k \bm{H}_k \bm{A} \bm{R}(t-1) \bm{A}' \bm{H}_k' \bm{T}_k' \nonumber\\
  &&  + [\bm{T}_k \bm{H}_k \bm{A} \hat{\bm{x}}(t-1)-\bm{T}_k \bm{H}_k \bm{A} \hat{\bm{x}}^{(k)}(t-1)+\bm{M}_k \bm{\theta}^{(k)}(t-1)+\bm{d}_k ] \nonumber\\
  && [\bm{T}_k \bm{H}_k \bm{A} \hat{\bm{x}}(t-1)-\bm{T}_k \bm{H}_k \bm{A} \hat{\bm{x}}^{(k)}(t-1)+\bm{M}_k \bm{\theta}^{(k)}(t-1)+\bm{d}_k ]' \bigg) \bm{\Sigma}_k^{-\frac{1}{2}} \bigg)
 \end{eqnarray}
 \normalsize
\hrule
\end{figure*}

Clearly, $\mathbb{E}(||\bm{\theta}^{(k)}(t)||^2| \mathcal{F}_{t-1})$  can be expressed as  \eqref{eqn:variance-of-theta}. 
Note that, given $\{\bm{\theta}^{(k)}(t-1): 1 \leq k \leq N\}$, the function $\sum_{k=1}^N \mathbb{E}(||\bm{\theta}^{(k)}(t)||^2| \mathcal{F}_{t-1} )$ is quadratic in $\{\bm{T}_k, \bm{U}_k, \bm{M}_k, \bm{d}_k\}_{1 \leq k \leq N}$. 

On the other hand, given $\mathcal{F}_{t-1}$, $\bm{x}(t-1) \sim \mathcal{N}(\hat{\bm{x}}(t-1), \bm{R}(t-1))$ where $(\hat{\bm{x}}(t-1), \bm{R}(t-1))$ can be computed by a standard Kalman filter. Now, 

\footnotesize
\begin{eqnarray*}
 \tilde{\bm{z}}_k(t) 
 &=& \bm{T}_k \bm{z}_k(t) + \bm{b}_k(t) \\
 &=& \bm{T}_k \bm{y}_k(t)-\bm{T}_k \bm{H}_k \bm{A} \hat{\bm{x}}^{(k)}(t-1)+ \bm{b}_k(t)\\
  &=& \bm{T}_k (\bm{H}_k \bm{x}(t)+\bm{v}_k(t))-\bm{T}_k \bm{H}_k \bm{A} \hat{\bm{x}}^{(k)}(t-1)+ \bm{b}_k(t)\\
    &=& \bm{T}_k \bm{H}_k \bm{A} \bm{x}(t-1)+ \bm{T}_k \bm{H}_k \bm{w}(t-1)+\bm{T}_k \bm{v}_k(t)\\
    &&-\bm{T}_k \bm{H}_k \bm{A} \hat{\bm{x}}^{(k)}(t-1)+ \bm{b}_k(t)
\end{eqnarray*}
\normalsize

which, given $\mathcal{F}_{t-1}$, is distributed as $\mathcal{N}(\bm{T}_k \bm{H}_k \bm{A} \hat{\bm{x}}(t-1)-\bm{T}_k \bm{H}_k \bm{A} \hat{\bm{x}}^{(k)}(t-1)+\bm{M}_k \bm{\theta}^{(k)}(t-1)+\bm{d}_k, \bm{T}_k \bm{H}_k \bm{Q} \bm{H}_k' \bm{T}_k'+ \bm{T}_k \bm{R}_k \bm{T}_k'+ \bm{T}_k \bm{H}_k \bm{A} \bm{R}(t-1) \bm{A}' \bm{H}_k' \bm{T}_k'+\bm{S}_k)$. Hence, $\mathbb{E}(\tilde{\bm{z}}_k(t)' \bm{\Sigma}_k^{-1} \tilde{\bm{z}}_k(t)|\mathcal{F}_{t-1})$ is given by \eqref{eqn:variance-of-z}.

$\sum_{k=1}^N \mathbb{E}(\tilde{\bm{z}}_k(t)' \bm{\Sigma}_k^{-1} \tilde{\bm{z}}_k(t)|\mathcal{F}_{t-1})$ is also  quadratic in 
$\{\bm{T}_k, \bm{U}_k, \bm{M}_k, \bm{d}_k\}_{1 \leq k \leq N}$. In case of  non-stationary  attack, these   results will hold w.r.t. $\{\bm{T}_k(t), \bm{U}_k(t), \bm{M}_k(t), \bm{d}_k(t)\}_{1 \leq k \leq N}$.

Hence,  the function $f_t(\{\bm{T}_k, \bm{U}_k, \bm{M}_k, \bm{d}_k\}_{1 \leq k \leq N}) \doteq \sum_{j=1}^N \mathbb{E}(||\bm{\theta}^{(j)}(t)||^2 + \lambda(t-1) \tilde{\bm{z}}_k(t)' \bm{\Sigma}_k^{-1} \tilde{\bm{z}}_k(t)| \mathcal{F}_{t-1} )$ is also quadratic in $\{\bm{T}_k, \bm{U}_k, \bm{M}_k, \bm{d}_k\}_{1 \leq k \leq N}$. We also note that, if $\bm{T}_k$ is fixed (e.g., the identity matrix), then the function $f_t(\{ \bm{U}_k, \bm{M}_k, \bm{d}_k\}_{1 \leq k \leq N})$ is convex in $\{\bm{U}_k, \bm{M}_k, \bm{d}_k\}_{1 \leq k \leq N}$.

\subsection{The attack design algorithm}
\label{subsection:attack-design-algorithm-known-model}
In this subsection, we propose an {\em optimal linear attack algorithm for distributed CPS} (OLAAD). The OLAAD algorithm involves two-timescale stochastic approximation \cite{borkar08stochastic-approximation-book}, which is basically a stochastic gradient descent algorithm with a noisy gradient estimate; \eqref{eqn:unconstrained-optimization-problem} is solved via SPSA in the faster timescale, and $\lambda$ is updated in the slower timescale. 

The algorithm requires three positive step size sequences $\{a(t)\}_{t \geq 0}$, $\{b(t)\}_{t \geq 0}$ and $\{c(t)\}_{t \geq 0}$ satisfying the following criteria: (i) $\sum_{t=0}^{\infty} a(t)=\sum_{t=0}^{\infty} b(t)=\infty$, 
(ii) $\sum_{t=0}^{\infty} a^2(t)<\infty, \sum_{t=0}^{\infty} b^2(t)<\infty$, 
(iii) $\lim_{t \rightarrow \infty} \frac{b(t)}{a(t)}=0$, (iv) $\lim_{t \rightarrow \infty} c(t)=0$, and (v) $\sum_{t=0}^{\infty} \frac{a^2 (t)}{c^2(t)}<\infty$. The first three conditions are standard requirements for two-timescale stochastic approximation. The fourth condition ensures that the gradient estimate is asymptotically unbiased, and the fifth condition is required for the convergence of SPSA.

\vspace{2mm}
\hrule
\noindent {\bf The OLAAD algorithm}
 \hrule
 \vspace{0.5mm}
 \noindent {\bf Input:} $\{a(t)\}_{t \geq 0}$, $\{b(t)\}_{t \geq 0}$, $\{c(t)\}_{t \geq 0}$, $\alpha$, $\eta$, $J$,

\noindent {\bf Initialization:} $\bm{T}_k(0), \bm{S}_k(0)=\bm{U}_k'(0) \bm{U}_k(0)$, $\bm{M}_k(0)$, $\bm{d}_k(0)$ for all $k \in \mathcal{N}$, $\lambda(0)$

\noindent {\bf For $t=1,2,3,\cdots$:}

\begin{enumerate}
\item For each $1 \leq  k \leq N$, the attacker  generates  random matrices $\bm{\Delta}^{(k)}(t)$, $\bm{\Gamma}^{(k)}(t)$, $\bm{\Pi}^{(k)}(t)$ and $\bm{\beta}^{(k)}(t)$ having same dimensions as $\bm{T}_k(t-1)$, $\bm{U}_k(t-1)$, $\bm{M}_k(t-1)$ and $\bm{d}_k(t-1)$ respectively,  whose entries are uniformly and independently chosen from the set $\{-1,1\}$.

\item The attacker computes $\bm{T}_k^+ \doteq \bm{T}_k(t-1) + c(t) \bm{\Delta}^{(k)}(t)$, $\bm{T}_k^- \doteq \bm{T}_k(t-1) - c(t) \bm{\Delta}^{(k)}(t)$, $\bm{U}_k^+ \doteq \bm{U}_k(t-1) + c(t) \bm{\Gamma}^{(k)}(t)$, 
$\bm{U}_k^- \doteq \bm{U}_k(t-1) - c(t) \bm{\Gamma}^{(k)}(t)$, 
$\bm{M}_k^+ \doteq \bm{M}_k(t-1) + c(t) \bm{\Pi}^{(k)}(t)$, 
$\bm{M}_k^- \doteq \bm{M}_k(t-1) - c(t) \bm{\Pi}^{(k)}(t)$, $\bm{d}_k^+ \doteq \bm{d}_k(t-1) + c(t) \bm{\beta}^{(k)}(t)$, $\bm{d}_k^- \doteq \bm{d}_k(t-1) - c(t) \bm{\beta}^{(k)}(t)$, for all $1 \leq k \leq N$. The matrices $\bm{S}_k^+ \doteq (\bm{U}_k^+)'\bm{U}_k^+$ and $\bm{S}_k^- \doteq (\bm{U}_k^-)'\bm{U}_k^-$ are computed.
\item The attacker computes:
\begin{eqnarray}
 \kappa_t^+ &\doteq& \sum_{j=1}^N \mathbb{E} \bigg( (||\bm{\theta}^{(j)}(t)||^2 + \lambda(t-1) \tilde{\bm{z}}_j(t)' \bm{\Sigma}_j^{-1} \tilde{\bm{z}}_j(t)\nonumber\\
 && | \mathcal{F}_{t-1}, \{\bm{T}_k^+, \bm{S}_k^+, \bm{M}_k^+, \bm{d}_k^+\}_{1 \leq k \leq N}   \bigg)
\end{eqnarray} 
using  \eqref{eqn:variance-of-theta} and \eqref{eqn:variance-of-z} under $\{\bm{T}_k^+, \bm{S}_k^+, \bm{M}_k^+, \bm{d}_k^+\}_{1 \leq k \leq N}$. The attacker computes $\kappa_t^-$ in a similar way using $\{\bm{T}_k^-, \bm{S}_k^-,  \bm{M}_k^-, \bm{d}_k^-\}_{1 \leq k \leq N}$.
\item The attacker updates each element $(i,j)$ of $\bm{T}_k(t-1)$, $\bm{U}_k(t-1)$, $\bm{M}_k(t-1)$ and $\bm{d}_k(t-1)$ for all $1 \leq k \leq N$ as follows:

\footnotesize
\begin{eqnarray}
 \bm{T}_k(t)(i,j) &=& \bm{T}_k(t-1)(i,j)-a(t) \times \frac{(\kappa_t^+ - \kappa_t^-)}{2 c(t) \bm{\Delta}_{(i,j)}^{(k)}(t)} \nonumber\\
  \bm{U}_k(t)(i,j) &=& \bm{U}_k(t-1)(i,j)-a(t) \times \frac{(\kappa_t^+ - \kappa_t^-)}{2 c(t) \bm{\Gamma}_{(i,j)}^{(k)}(t)}\nonumber\\
  \bm{M}_k(t)(i,j) &=& \bm{M}_k(t-1)(i,j)-a(t) \times \frac{(\kappa_t^+ - \kappa_t^-)}{2 c(t) \bm{\Pi}_{(i,j)}^{(k)}(t)}\nonumber\\
  \bm{d}_k(t)(i,1) &=& \bm{d}_k(t-1)(i,1)-a(t) \times \frac{(\kappa_t^+ - \kappa_t^-)}{2 c(t) \bm{\beta}_{(i,1)}^{(k)}(t)}\nonumber\\
\end{eqnarray}
\normalsize

The attacker computes $\bm{S}_k(t) \doteq (\bm{U}_k(t))'\bm{U}_k(t)$ for all $1 \leq k \leq N$. 
\item The sensors make observations $\{\bm{y}_k(t)\}_{1 \leq k \leq N}$, which are accessed by the attacker.
\item The attacker calculates $\bm{z}_k(t)=\bm{y}_k(t)-\bm{H}_k \bm{A} \hat{\bm{x}}^{(k)}(t-1)$  for all $k \in \{1,2,\cdots,N\}$.
\item The attacker calculates  $\bm{\tilde{z}_k}(t)=\bm{T}_k(t) \bm{z}_k(t)+\bm{b}_k(t)$ for all $k \in \{1,2,\cdots,N\}$, where $\bm{b}_k(t) \sim \mathcal{N}(\bm{M}_k(t) \bm{\theta}^{(k)}(t-1)+\bm{d}_k(t), \bm{S}_k(t))$ chosen independently of all other variables. The observations are accordingly modified as $\tilde{\bm{y}}_k(t)=\tilde{\bm{z}}_k(t)+\bm{H}_k \bm{A} \hat{\bm{x}}^{(k)}(t-1)$ and sent to the agent nodes.
\item The attacker updates the Lagrange multiplier as follows: 
\begin{equation}
 \lambda(t)=\lambda(t-1)+b(t) (\sum_{k=1}^N \tilde{\bm{z}}_k(t)' \bm{\Sigma}_k^{-1} \tilde{\bm{z}}_k(t)-\frac{\alpha \eta}{J})
\end{equation}

\item The agent nodes compute the estimates locally, using \eqref{eqn:KCF-equation} and  the modified $\{\tilde{\bm{y}}_k(t)\}_{1 \leq k \leq N}$. The agent nodes broadcast their estimates to their neighbouring nodes.
\end{enumerate}
\noindent {\bf end}
\label{algorithm:correction-algorithm-learning} 
 \hrule
\vspace{2mm}

Note that, if $\{\bm{T}_k\}_{1 \leq k \leq N}$ is kept fixed, then the first update in step~$4$ of OLAAD is not required. 

The OLAAD algorithm combines the online stochastic gradient descent (OSGD)  algorithm of \cite[Chapter~$3$]{hazan2016introduction} with two-timescale stochastic approximation of \cite{borkar08stochastic-approximation-book}. The $\lambda(t)$ iterate is updated in the slower timescale to meet the constraint in \eqref{eqn:constrained-optimization-problem}. In the faster timescale, OSGD is used for solving  \eqref{eqn:unconstrained-optimization-problem}. Since $\lim_{t \rightarrow \infty} \frac{b(t)}{a(t)}=0$, the faster timescale iterates $\{\bm{T}_k(t), \bm{U}_k(t), \bm{M}_k(t), \bm{d}_k(t)\}_{1 \leq k \leq N}$ view the slower timescale iterate $\lambda(t)$ as quasi-static, while the $\lambda(t)$ iteration finds the faster timescale iterates as almost equilibriated; as if, the faster timescale iterates are varied in an inner loop and the slower timescale iterate is varied in an outer loop. Note that, OLAAD has no guarantee of convergence to globally optimal solution in general.

\section{Numerical results}\label{section:numerical-work}
\subsection{Three agent nodes, scalar proces}
We consider a line topology with $N=3$ agent nodes. Process dimension $q=1$ and the observation dimension at each node is $2$.  The system parameters $\bm{A}, \bm{Q}, \{\bm{R}_k\}_{1 \leq k \leq 3}, \{\bm{H}_k\}_{1 \leq k \leq 3}$ are chosen randomly. The KCF parameters $\{\bm{G}_k, \bm{C}_k\}_{1 \leq k \leq 3}$ are computed using a technique from \cite{saber09kalman-consensus-optimality-stability}, and $\{\bm{\Sigma}_k\}_{1 \leq k \leq 3}$ are computed by simulating the KCF under no attack. 

For FDI attack, we set  $\bm{x}^*=5$, $\alpha=0.3$, $\eta=500$, $J=3$, $\lambda(0)=7$, $a(t)=\frac{0.5}{t^{0.6}}$, $b(t)=\frac{0.5}{t^{0.9}}$, $c(t)=\frac{0.01}{t^{0.1}}$, $\bm{T}_k(t)=\bm{I}$ for all $1 \leq k \leq 3$ and for all  $t \geq 1$,  simulated the performance of OLAAD for $1000000$ iterations and evaluated its performance between $500000$-th to $1000000$-th iteration. However, motivated by the ADAM algorithm from \cite{KingmaB14}, we implemented an adaptive step size version of SGD with the basic step sizes being $\{a(t), b(t)\}_{t \geq 1}$,\footnote{Note that, under ADAM, the conditions on step sizes in Section~\ref{subsection:attack-design-algorithm-known-model} are not necessarily satisfied for all sample paths.} though gradient estimation was done via simultaneous perturbation using step size $\{c(t)\}_{t \geq 1}$. In all problem instances, deviation from $\bm{x}^*$ means the squared distance of $\bm{x}^{(k)}(t)$ from $\bm{x}^*$, summed over nodes and averaged over time; similar definition applies to deviation from origin.

We simulated multiple problem instances for different sample paths; some results are tabulated below:

\begin{table}[h!]
\footnotesize
\centering
\begin{tabular}{|c |c |c |c |c |c |}
\hline
  Problem & Attack  & Deviation & Deviation   & Deviation & Deviation \\
 instance &   detection & from $\bm{x}^*$  & from $\bm{x}^*$ & from origin & from origin \\ 
 &probability &  under FDI & (no attack) & under FDI & (no attack) \\
 \hline
1 & 0 & 29.3367 & 75.3063 & 11.0587 &  0.3215\\
2 & 0  & 27.6588 & 75.4367 & 12.2885 & 0.4373\\
3  & 0 & 42.7055 & 75.3845 & 5.0898 & 0.4001\\
\hline
 \end{tabular}
\normalsize
\end{table}

We notice that the attack detection probability is $0$ because we consider a stronger constraint (upper bound to the actual attack detection probability) in our formulation. The attack detection probability also depends on the system realization, and the values of $\eta$ and $J$. For some other values of $\eta$ and $J$ and various system realizations, we obtained the following:

\begin{table}[h!]
\footnotesize
\centering
\begin{tabular}{|c |c |c |c |c |c |}
\hline
  Problem & Attack  & Deviation & Deviation   & Deviation & Deviation \\
 instance &   detection & from $\bm{x}^*$  & from $\bm{x}^*$ & from origin & from origin \\ 
 &probability &  under FDI & (no attack) & under FDI & (no attack) \\
 \hline
1 & 0.1190 & 70.8189 & 75.0383 & 0.2398 & 0.0014 \\
2 & 0.0047  & 46.1039 &  75.2841 & 11.7185 & 0.3227\\
\hline
 \end{tabular}
\normalsize
\caption{$N=3, q=1$}
\label{table:table1}
\end{table}

 We observe that OLAAD significantly reduces the deviation from $\bm{x}^*=5$, and increases the deviation from the origin which is the mean of $\bm{x}(t)$. In all instances, constraint on attack detection probability is satisfied. This shows that OLAAD is a viable FDI attack scheme for distributed CPS.

\subsection{Five agent nodes, vector process} 
Here we used the same setting as before, except that $N=5$, process dimension $q=2$, $\bm{x^*}=[5 5]^T$, $\alpha=0.1$, $\eta=200$, $J=2$, $a(t)=\frac{0.05}{t^{0.6}}$, $b(t)=\frac{0.05}{t^{0.9}}$. The results are tabulated next. It is to be noted that here also we  observe the same pattern in results as seen for $N=3$. 

\begin{table}[h!]
\footnotesize
\centering
\begin{tabular}{|c |c |c |c |c |c |}
\hline
  Problem & Attack  & Deviation & Deviation   & Deviation & Deviation \\
 instance &   detection & from $\bm{x}^*$  & from $\bm{x}^*$ & from origin & from origin \\ 
 &probability &  under FDI & (no attack) & under FDI & (no attack) \\
 \hline
1 & 0.0025 & 29.0400  &  251.3088 & 135.2330  & 1.3124  \\
2 & 0  &  145.9975 & 252.0693  & 42.1946  & 2.0549 \\
3  & $8.5 \times 10^{-5}$  & 22.6360  & 255.0649  &  177.5157 & 5.1584 \\
\hline
 \end{tabular}
\normalsize
\end{table}

However, it was also noticed across problem instances that, if the initial step size values $a(1), b(1) , c(1)$ are significant, then the iterated might be unstable due to high initial fluctuation; hence, the initial values of the step sizes need to be chosen carefully.

\section{Conclusions}\label{section:conclusion}
In this paper, we have designed an optimal linear attack for distributed cyber-physical systems. The parameters of the attack scheme were learnt and optimized on-line. Numerical results demonstrated the efficacy of the proposed attack scheme. In future, we seek to extend this work for unknown process and observation dynamics and also prove convergence of the proposed algorithms.

{\small
\bibliographystyle{unsrt}
\bibliography{arpantechreport}
}

%
%

%
%

\end{document}